\documentstyle[aps,twocolumn,epsfig]{revtex}
\begin{document}
\newcommand{\ve}[1]{\mbox{\boldmath $#1$}}
\twocolumn[\hsize\textwidth\columnwidth\hsize
\csname@twocolumnfalse%
\endcsname
 
\draft

\title {Sound Propagation in Elongated Bose-Einstein Condensed Clouds}
\author{G. M. Kavoulakis$^1$ and C. J. Pethick$^{1,2}$}
\date{\today} 
\address{$^1$Nordita, Blegdamsvej 17, DK-2100 Copenhagen \O, Denmark, \\
        $^2$Department of Physics, University of Illinois at
        Urbana-Champaign, 1110 W. Green Street, Urbana, IL 61801-3080}
\maketitle
 
\begin{abstract}

     We consider propagation of sound pulses along the long axis of a
Bose-Einstein condensed cloud in a very anisotropic trap. In the linear
regime, we demonstrate that the square of the velocity of propagation is
given by the square of the local sound velocity, $c^2=nU_0/m$, averaged
over the cross section of the cloud. We also carry out calculations in
the nonlinear regime, and determine how the speed of the pulse depends
on its amplitude.

\end{abstract}
\pacs{PACS numbers: 03.75.Fi, 05.30.Jp, 67.40.Db}
 
\vskip2pc]

   In a beautiful recent experiment, Andrews {\it et al.} have reported 
measurements of the propagation of
sound pulses along the axis of an elongated cloud of Bose-Einstein condensed
Na atoms \cite{andrews}.  At the center of the trap a disturbance was 
created by turning a laser on or off.  Due to the energy shift caused by
the interaction of atoms with the laser field, this led to a 
change in the potential experienced by 
atoms, and thereby to a density disturbance, whose time development was 
studied by phase-contrast imaging. In a dilute, homogeneous Bose gas at zero 
temperature, the speed of sound, $c_0$, is given by 
$c_0^2=dP/d\rho=n_0U_0/m$, 
where $P$ is the pressure, $\rho =mn$ is the mass density, $m$ is the 
atomic mass, $n$ is the density of atoms, and 
$U_0=4\pi \hbar^2 a/m$ is the strength of 
the effective two-body interaction, with $a$ being the  scattering length
for atom-atom collisions \cite{bogoliubov,lee}.  The experimental data 
seemed to be in accord with this result, if the 
density was taken to be that at the center 
of the cloud.  Within the accuracy of the measurements, no evidence for a 
dependence of propagation speed on amplitude was seen. 

   The purpose of this Letter is to investigate how propagation of a sound 
pulse is influenced by the inhomogeneity of the cloud, 
and by nonlinear effects.  The important result of our 
calculations is that, in the linear regime, the sound speed is given
by the expression for the homogeneous gas, but evaluated for the average
density over the cross section of the cloud perpendicular to the direction 
of propagation.  A second result is that nonlinear effects are substantial, 
and can give rise to changes in the velocity of the pulse of order of 10\%, for 
relative density perturbations of order 15\%.
 
   Consider a Bose-Einstein condensed cloud whose transverse dimensions ($\sim
R_{\perp}$) are much less than the axial dimension ($\sim Z$). We study the
propagation along the axis of the trap, of pulses whose characteristic
spatial scale, $l$, in the axial direction is large compared with $R_{\perp}$,
but small compared with $Z$.  That this condition is satisfied in the 
experiments may easily be seen by inspection of the density profiles in Fig.\,2
of Ref. \cite{andrews}. For the moment we make no detailed assumptions about
the confining potential perpendicular to the long direction of the trap,  
which we take to be the $z$-axis, apart from the fact that
its variations with respect to $z$ should be small on a length scale $l$.  
The fact that the transverse
dimensions of the cloud are small compared with the scale of the pulse,
means that the density profile across the trap may be assumed to have its
equilibrium form appropriate to the local number of particles per unit
length. This follows from the fact that the timescale for adjustment of 
the profile ($\sim
R_{\perp}/c$), where $c$ is a typical sound speed, is short compared with
the time for passage of the pulse, $\sim l/c$. The problem then becomes a 
one-dimensional one, and the sound pulse may be specified in terms of a
local velocity, $v(z)$, and a local density of particles, $\sigma(z)$, per
unit length.  The latter is given in 
terms of the particle density, $n(x, y, z)$, where $x$ and $y$
are coordinates perpendicular to the axis of the trap, by
\begin{eqnarray}
     \sigma(z)=\int dx dy \, n(x,y,z).
\label{sigma}
\end{eqnarray}

    The equations governing the motion of the condensate may be derived from 
the Gross-Pitaevskii equation, and they are just the equations of perfect 
fluid hydrodynamics, with the addition of a ``quantum pressure" term in the 
Euler equation \cite{GP}. The quantum pressure term 
is important only on length scales of order the coherence length, $\xi
=(8\pi n a)^{-1/2}$, which is typically of order 0.2-0.4 $\mu$m at the center
of the cloud \cite{andrews}. This is  considerably
less than the lengths of the pulses in the experiment,
which are of order 10 $\mu$m, and therefore we shall neglect the quantum
pressure term in these initial studies. However, this term will be
important when shocks or other sharp structures develop.
The equation of motion for $\sigma$ is obtained from the
continuity equation by integrating over the cross
section of the cloud, and is
\begin{eqnarray}
    \frac{\partial \sigma}{\partial t}  +  \frac{\partial
  (\sigma v)}{\partial z}=0.
\label{cont}
\end{eqnarray}
The equation of motion for $v$ is found by integrating the Euler equation over 
the cross section of the cloud, 
\begin{eqnarray}
    m \sigma \frac{dv}{dt}=-\int dx dy \,\frac{\partial
  P}{\partial z} -\sigma \frac{\partial V}{\partial z},
\label{nav-sto}
\end{eqnarray}
where $V$ is the external potential due to the trap and the interaction
with the laser field. In deriving this result, we 
have assumed that the dependence of $V$ on $x$
and $y$ may be neglected. We shall assume that the number of particles is 
so large that the Thomas-Fermi approximation is valid \cite{bp}. For
a spatially-uniform, dilute Bose gas at zero temperature, the energy per
unit volume (and also the pressure) are given by
$E=n^2 U_0/2$. The change in the local pressure is thus given by $dP 
=nU_0 dn$, where $dn$ is the local change in density.
 The chemical potential, $\mu$, neglecting the effects of the 
trapping potential, is $n U_0$, and in the Thomas-Fermi approximation 
the sum of this and the trap potential
$V$ is a constant for a given value of
$z$.  Thus changes in the chemical potential must likewise be functions
only of $z$, and, since
$d\mu=U_0dn$, it follows that changes in the particle density are
independent of $x$ and $y$.  The change in density is thus given simply in
terms of the change in the number of particles per unit length by
$dn=d\sigma/A$, where $A$ is the cross-sectional area of the cloud, and 
therefore the Euler equation becomes
\begin{eqnarray}
    m\sigma \frac{dv}{dt} &=& -U_0\int dx dy \,n
  \frac{\partial n}{\partial z} -\sigma \frac{\partial V}{\partial z} 
\nonumber \\ &=& 
 -U_0\frac{\sigma}{A}\frac{\partial
 \sigma}{\partial z}-\sigma \frac{\partial V}{\partial z}.
\label{euler}
\end{eqnarray}

     Let us begin by considering the linear regime. If we assume that the 
variation of the potential in the $z$-direction may be neglected, the 
problem becomes translationally invariant in this direction, and  from the 
linearized versions of the continuity equation, Eq.\,(\ref{cont}), 
and the Euler equation, Eq.\,(\ref{euler}), we find 
\begin{eqnarray}
     \frac{\partial^2\sigma}{\partial t^2}-
    \frac{\bar{n} U_0}{m}\frac{\partial^2 \sigma}{\partial z^2}=0,
\label{nav-sto3}
\end{eqnarray}
where $\bar{n}=\sigma/A$ is the mean density over the cross section of the
cloud.   This result demonstrates that the pulse satisfies the wave
equation with a velocity, $c(z)$, given by
\begin{eqnarray}
     c^2(z)=\frac{\bar{n} U_0}{m}.
\label{vel}
\end{eqnarray}
For the case of confinement by a harmonic trap potential in the transverse 
directions, this result has been derived independently by Zaremba 
\cite{zaremba} using different methods.  In this case $\bar n=n(\rho=0)/2$, 
where $n(\rho=0)$ is the density on the axis of the trap.  Here $\rho$ is 
the radius in cylindrical polar coordinates.  In the presence of a 
confining potential along the axis of the trap, Eq.\,(\ref{nav-sto3})
with the average density replaced by its local value, $n(\rho=0, z)$, will 
be a good approximation, provided the characteristic length of the 
disturbance is small compared with the axial dimension of the cloud.

   The above derivation underlines the generality of the result for the sound 
speed, but it is instructive to consider an alternative approach for the
specific case of a harmonic confining potential in the transverse
direction.  The basic idea is to treat the system as a one-dimensional
``elastic'' medium.  The energy of the system consists of the   kinetic 
energy due to the bulk motion of the condensate, an ``elastic" energy that 
takes into account contributions from particle interactions and the trap 
potential in the transverse direction, and a contribution  due to the 
$z$-component of the trap potential.
For simplicity we shall assume that the confining potential in
the transverse direction is given by $U(\rho)=m\omega^2\rho^2/2$, where
$\omega$ is the frequency of transverse oscillations of a single particle
in the trap. The kinetic energy, $T$, per unit length is
\begin{eqnarray}
      T=\frac{1}{2}\int dz \,\sigma
   \left(\frac{\partial\zeta}{\partial t}\right)^2,
\end{eqnarray}
where $\zeta(z)$ is the displacement of a particle of the fluid from its
equilibrium position. We next evaluate the elastic energy, which may be 
calculated assuming the cross section of the cloud to be locally 
uniform. Per unit length, the energy due to the dependence of the trap 
potential on the transverse coordinates is given by
\begin{eqnarray}
	E_{\rm trap}= \frac{1}{2} m\omega^2\int 2 \pi \rho d\rho 
  \, n(z, \rho) \rho^2,
\label{trap}
\end{eqnarray}
where the integral is to be taken over the cross section of the cloud.
The interaction energy per unit volume is given by
\begin{eqnarray}
	E_{\rm int} =\frac{1}{2}U_0\int 2 \pi \rho d\rho \, n^2(z, \rho).
\label{int}
\end{eqnarray}
It is  convenient to express quantities in terms of the
number of particles per unit length. For the harmonic oscillator trap, the
density profile in Thomas-Fermi theory is given by \cite{bp}
\begin{eqnarray}
    n(\rho,z)=n(0, z)\left(1-\frac{\rho^2}{R^2(z)}  \right),
\label{rho}
\end{eqnarray}
where $R(z)$ is the radius of the cross section
of the cloud. Thus the number of particles per unit length is given by
\begin{eqnarray}
	\sigma(z)=\frac{1}{2}n(0,z)\pi R^2.
\label{sigmaz}
\end{eqnarray}
The  density on the axis of the trap is given by
\begin{eqnarray}
    n(0,z)U_0=\frac{1}{2}m\omega^2 R^2,
\label{condition}
\end{eqnarray}
and thus one finds that both $n(0, z)$ and $R^2$ vary as $\sigma^{1/2}$.
On performing the integrations in Eqs.\,(\ref{trap}) and (\ref{int}), one
arrives at the following expression for the elastic energy per unit length:
\begin{eqnarray}
	E_{\rm el}=\frac{\pi}{3}U_0 n^2(0,z) R^2 =\frac{2}{3} 
\left(\frac{m\omega^2 U_0}{\pi}\right)^{1/2} \sigma^{3/2}.
\label{elastic}
\end{eqnarray}
Note that this increases less rapidly with increasing particle number per
unit length than it would in the case of confinement by a rigid pipe with
infinitely high walls, in which case the potential energy scales as
$\sigma^2$.  This reflects the fact that for a harmonic confining potential
an increase in the number of particles per unit length results in expansion
of the cloud in the radial direction, thereby leading to a less rapid
increase of the total potential energy than would be the case for
confinement by rigid walls.  

    Let us now use the result of Eq.\,(\ref{elastic}) to calculate the velocity 
of sound in a Bose-condensed cloud when there is no potential acting along the 
axis. The force parallel to the axis of the trap is given by $F=   
\sigma^2d(E_{\rm el}/\sigma)/d\sigma$, and the sound speed is given by  
$mc^2=dF/d\sigma =\sigma d^2E_{\rm el}/d\sigma^2$, results completely
analogous to those for the pressure and the sound speed in a bulk medium.  
Inserting the expression Eq.\,(\ref{elastic}) into this expression, one finds
$c^2=n(\rho=0)U_0/(2m)$.  Since for a trap which is harmonic in the transverse 
directions the average density is half the density on the axis, this result 
agrees with Eq.\,(\ref{vel}).  

   We now consider how a pulse will propagate along the axis of the trap when 
allowance is made for the varying density.  In the Thomas-Fermi 
approximation, the local density on the axis of the trap is given by  $n(0, 
z) =n_{\rm max}(1-z^2/Z^2)$, where $n_{\rm max}$ is the density at the 
center of the cloud, and $Z$ is the axial dimension of the cloud.  Thus the 
local sound speed varies as $(1-z^2/Z^2)^{1/2}$. In the WKB approximation a 
disturbance initially at $z=0$ will at time $t$ have propagated a 
distance $z$ determined by 
\begin{eqnarray}
    t(z)=\int_0^z \frac {dz} {c(z)}=\frac {Z} {c(0)}
\sin^{-1} \frac z Z,
\label{timeee}
\end{eqnarray}
or $z=Z\sin[c(0) t/Z]$, where $c(0) =[n_{\rm 
max}U_0/(2m)]^{1/2}$ is the sound velocity, Eq.\,(\ref{vel}), at $z=0$. 

   In the experiments \cite{andrews}, relative density disturbances 
were as large as 100\%, so 
it is of interest to investigate nonlinear effects.  The basic equations 
are Eqs.\,(\ref{cont}) and (\ref{euler}), which are a closed set of 
equations for $v$ and $\sigma$, since from Eqs.\,(\ref{sigmaz}) and 
(\ref{condition}) it follows that the area $A$ is given by $A=2[\pi \sigma 
U_0/(m \omega^2)]^{1/2}$.  We have integrated these equations by 
employing the method of characteristics \cite{landau} for a number of cases 
that correspond to the experimental conditions.  In one set of calculations 
the cloud was initially in equilibrium in a potential due to the trap and an 
extra repulsive 
potential representing the interaction of atoms with the radiation field of 
the laser.  The extra potential was taken to have the 
form $V_{\rm rad}=V_0\exp({-2z^2/l^2})$, where $l=12$ $\mu$m, as in the 
experiments, and we took the axial extent of the condensate, $2Z$, to be 
450 $\mu$m.  At time $t=0$ the repulsive potential was turned off.  In other 
runs we started with a configuration in equilibrium 
in the presence of the trap alone, and turned the laser 
field on at $t=0$.  The strength of the potential due to the laser could 
be varied to produce different perturbations in the density.

\begin{figure}
\begin{center}
\epsfig{file=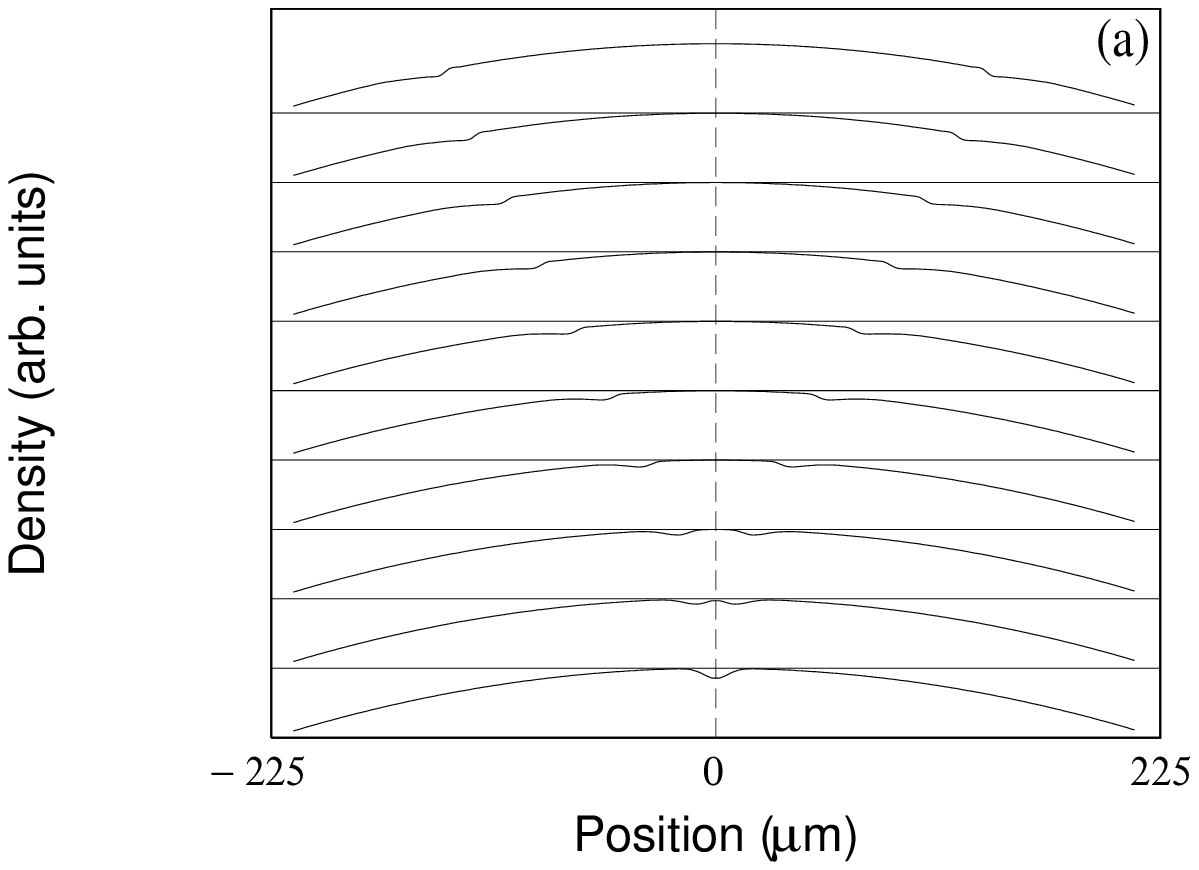,width=\linewidth,height=7cm}
\vskip1pc
\epsfig{file=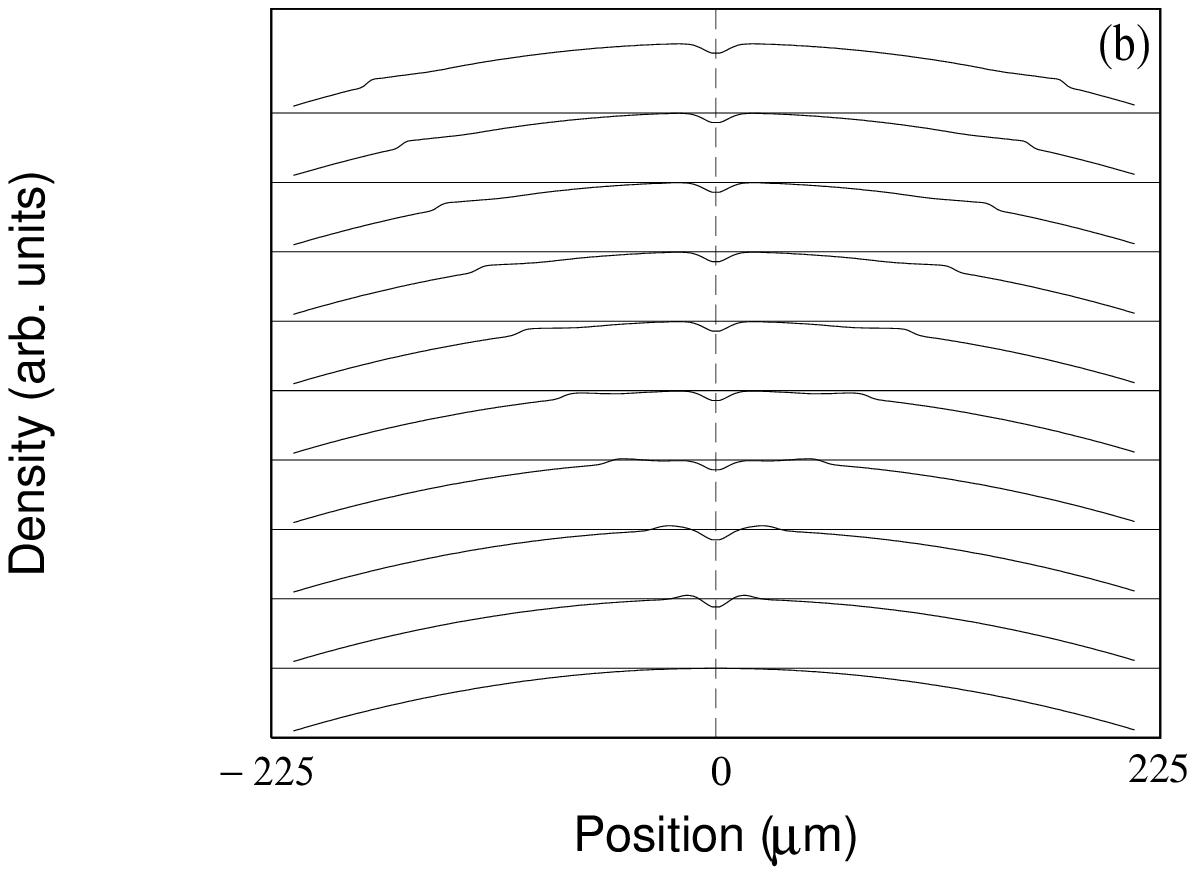,width=\linewidth,height=7cm}
\begin{caption}
{Density profiles of the gas along the $z$-axis for various times
for cases when the perturbing potential is turned off, (a), or turned on (b).
The profiles are shown for times 0, 3, 6, 12, 18, 24, 30,
36, 42 and 48 ms.}
\end{caption}
\end{center}
\label{FIG1}
\end{figure}
   In Fig.\,1 we show density profiles along the $z$-axis of the cloud for
various times for a laser intensity that, in equilibrium, would create a
relative density perturbation of $15\%$ at the center of the cloud.
We observe that
positive density pulses travel faster than negative ones, as one would expect 
from the fact that the sound speed increases with density. In Fig.\,2 we plot 
the positions of extrema of the pulses as functions of time. For reference we 
also show corresponding results for low amplitudes, which were obtained by 
taking the average of the results for switching off a potential having the same 
spatial dependence, but whose strength was such as to create a relative density 
perturbation of either $+5 \%$, or $-5\%$. The agreement with the result of the 
linear theory, Eq.\,(\ref{timeee}), which is shown as a solid line, is very  
good, and the small difference compared with the numerical results is
probably due to the varying cross section of the cloud.
Note that pulses become distorted as they propagate.  This is a consequence 
of nonlinear effects, and the distance a pulse can travel
before it is distorted 
significantly may be estimated by observing that the sound velocity varies as 
$n^{1/2}$, where $n$ is the density, and therefore the sound velocity change 
associated with a density change $\delta n$ is approximately $(c/2)\delta n/n$. 
In a time $t$ a disturbance in the region at density $n+\delta n$ will move a 
distance $\sim (c t/2)\delta n/n$ relative to what it would if the density were 
$n$, and consequently distortion of the pulse will be appreciable if this 
distance is greater than or of order the length of the pulse, $l$.  In terms of 
the distance travelled by the pulse, $z\approx ct$, the condition for there to 
be significant distortion is $z \, {\raisebox{-.5ex}{$\stackrel{>}{\sim}$}}\,
2 \, l \,n/\delta n$, an estimate which appears to be in accord with the 
numerical calculations.
\begin{figure}
\begin{center}
\epsfig{file=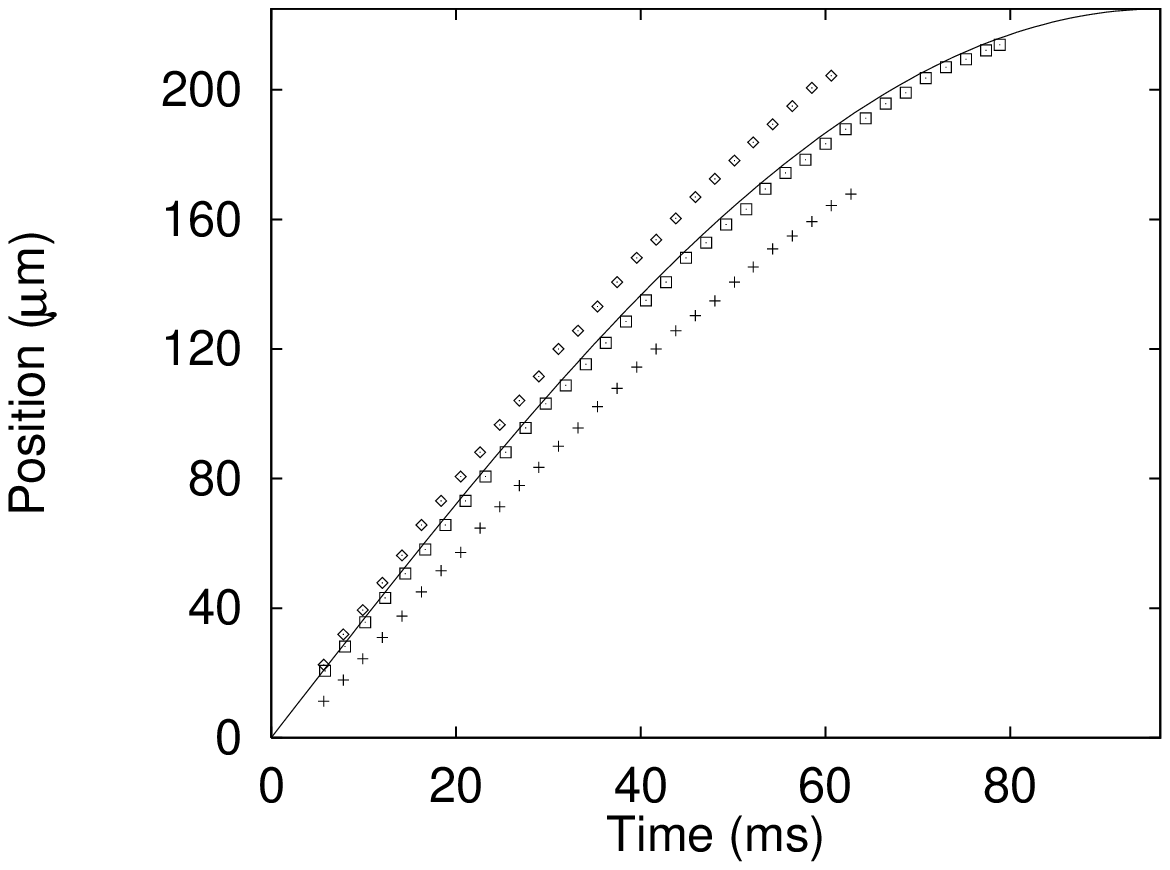,width=\linewidth,height=6.5cm}
\begin{caption}
{Distance of the extrema of the density from the center of the cloud
as a function of time. Squares give results for low amplitude pulses, as
explained in the text, and the solid line is the analytical result,
Eq.\,(\ref{timeee}). Diamonds give results for the case shown in Fig.\,1a,
while crosses correspond to that shown in Fig.\,1b.
The central density is taken to be $10^{14}$ cm$^{-3}$,
which corresponds to a time $\pi Z/[2c(0)] =96.3$ ms for a linear pulse to
reach the end of the cloud.}
\end{caption}
\end{center}
\label{FIG2}
\end{figure}

     To exhibit stronger nonlinear effects, we show in Fig.\,3,
results similar to those in Fig.\,1 but for
a perturbing potential that would create a $30\%$ density perturbation in 
equilibrium.  The steepening of the leading edge of positive density 
perturbations and of the trailing edge of negative ones is apparent.  At times 
later than those for which results are shown, shocks develop. To understand 
later stages of the evolution, it is necessary to develop a basic understanding 
of the physics of shocks in superfluids.  In the MIT experiments density 
perturbations were as large as 100\%, and under such conditions  
nonlinear effects are expected to be extremely strong.  An important future 
experimental challenge is to put in evidence the predicted effects on sound 
propagation of nonuniformity of the particle density over the cross section of 
the trap, and of nonlinearity.  Our results indicate that, because of their 
high compressibilities, low density atomic 
Bose-Einstein condensates are useful systems for investigation of nonlinear 
phenomena.
\begin{figure}
\begin{center}
\epsfig{file=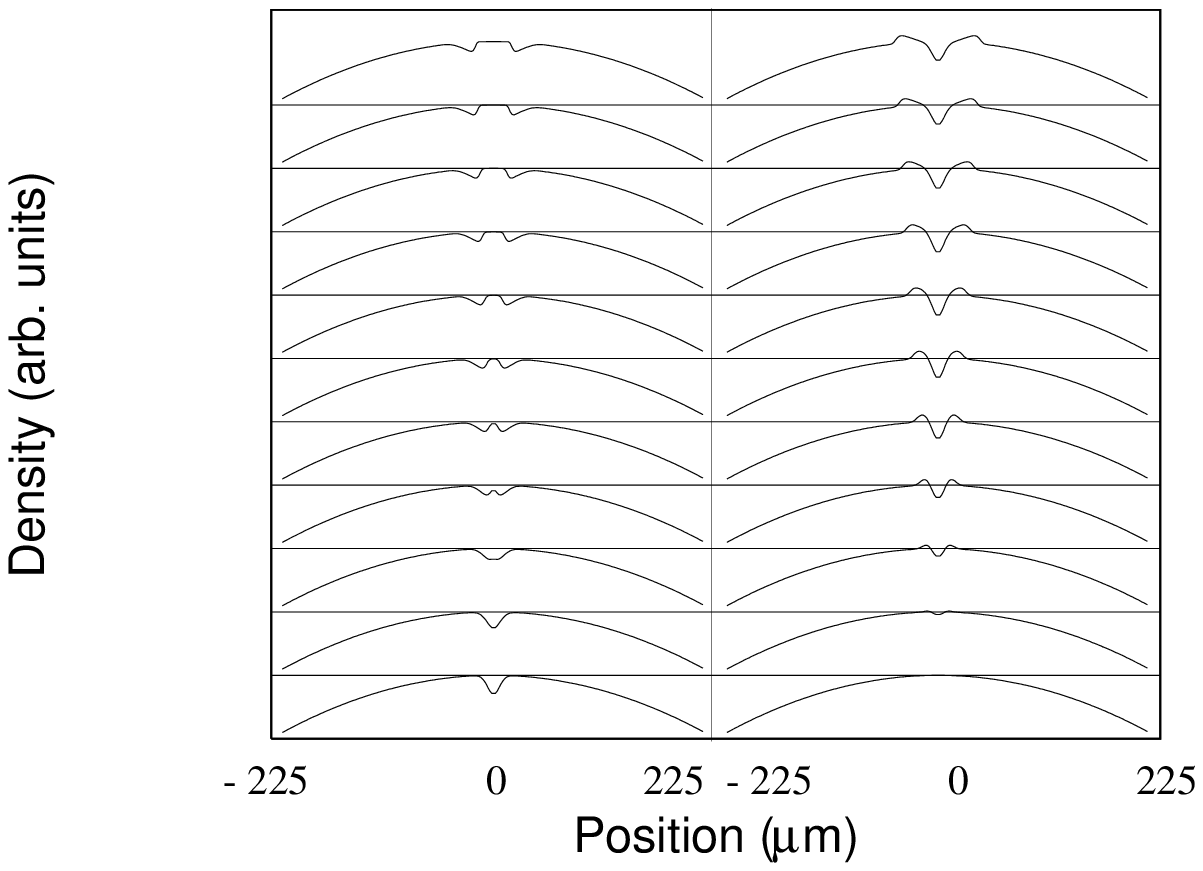,width=\linewidth,height=7cm}
\begin{caption}
{Density profiles of the gas along the $z$-axis for various times
for a larger density perturbation. 
The situations correspond to those in Figs.\,1a and 1b,
apart from the fact that the strength of the laser field was chosen
so that in equilibrium it would create a $30\%$ density perturbation.  The
profiles are shown for times equal to multiples of 0.85 ms, beginning at $t=0$.}
\end{caption}
\end{center}
\label{FIG3}
\end{figure}

     Helpful discussions with K. Andersen, A. D. Jackson,
W. Ketterle, D. M. Kurn, H.-J. Miesner, \AA. Nordlund, and H. Smith
are gratefully acknowledged. One of us (G.M.K.) would particularly like
to thank G. Baym, who introduced him to the method of characteristics in
connection with an earlier problem. G.M.K. would also like to thank the
Foundation of Research and Technology, Hellas (FORTH) for its hospitality.

\end{document}